\documentclass{aa}  

\usepackage{graphicx}
\usepackage{txfonts}
\usepackage{graphicx}
\graphicspath{ {./plots/} }
\usepackage[colorlinks,linkcolor=blue,citecolor=blue]{hyperref}
\usepackage{placeins}
\usepackage{graphicx}
\usepackage{xcolor}
\usepackage{siunitx}
\usepackage{xspace}
\usepackage{natbib}

\newcommand{\rc}{\ensuremath{r_\mathrm{c}}\xspace}
\usepackage[version=4]{mhchem} 
\usepackage{float}

\begin{document}

\renewcommand{\figureautorefname}{Fig.}
\renewcommand{\sectionautorefname}{Section}
\renewcommand{\subsectionautorefname}{Section}
\renewcommand{\equationautorefname}{Eq.}

\title{A tension between dust and gas radii: the role of substructures and external photoevaporation in protoplanetary disks}

\author{Luca Delussu
    \inst{1},
    Rossella Anania\inst{2},
    Tilman Birnstiel\inst{1,3},
    Claudia Toci \inst{4,5},
    Giovanni Rosotti\inst{2},
    Sebastian Markus Stammler\inst{1},
    Tommy Chi Ho Lau\inst{6,1}, \and
    Anna Miotello\inst{4}
}

\institute{University Observatory, Faculty of Physics, Ludwig-Maximilians-Universität München, Scheinerstr. 1, 81679 Munich, Germany
    \and Dipartimento di Fisica ’Aldo Pontremoli’, Università degli Studi di Milano, via G. Celoria 16, I-20133 Milano, Italy
    \and
    Exzellenzcluster ORIGINS, Boltzmannstr. 2, D-85748 Garching, Germany
    \and European Southern Observatory, Karl-Schwarzschild-Strasse 2, 85748 Garching bei München, Germany
    \and INAF, Osservatorio Astrofisico di Arcetri, 50125 Firenze, Italy
    \and Department of Astronomy and Astrophysics, University of Chicago, Chicago, 60637 IL, USA}

\abstract
{Protoplanetary disk substructures are thought to play a crucial role in disk evolution and planet formation. Population studies of disks large-sample size surveys show that substructures, and their rapid formation, are needed to reproduce the observed spectral indices. Moreover, they enable the simultaneous reproduction of the observed spectral index and size-luminosity distributions.}
{This study aims to investigate the necessity of substructures and predict their characteristics to reproduce gas-to-dust size ratios observed in the Lupus star-forming region.}
{We performed a population synthesis study of gas and dust evolution in disks using a two-population model (two-pop-py) and the \texttt{DustPy} code. We considered the effects of viscous evolution, dust growth, fragmentation, transport, and external photoevaporation. The simulated population distributions were obtained by post-processing the resulting disk profiles of surface density, maximum grain size, and disk temperature.}
{Although substructures help reduce the discrepancy between simulated and observed disk gas-to-dust size ratios; even when accounting for external photoevaporation, they do not fully resolve it. Only specific initial conditions in disks undergoing viscous evolution with external photoevaporation can reproduce the observations, highlighting a fine-tuning problem. While substructured disks reproduce dust size and spectral index, they tend to overestimate gas radii.}
{The results ultimately highlight the main challenge of simultaneously reproducing gas and dust sizes. One possible explanation is that the outermost substructure is linked to the disk truncation radius, which determines the gas radius, or that substructures are frequent enough to always be near the gas outer radius.}

\titlerunning{short title}
\authorrunning{short title}
\maketitle

\section{Introduction}

The study and interest in protoplanetary disk substructures have been significantly fueled by the advent of the Atacama Large Millimeter/Sub-Millimeter Array (ALMA) and its high-resolution observations. The iconic image of HL Tau, captured by \cite{brogan20152014}, not only provided the first direct image of a protoplanetary disk but also revealed, for the first time, the presence of substructures. In particular, surveys such as the Disk Substructures at High Angular Resolution Project (DSHARP) \citep{Andrews2018, Huang2018}, \cite{Long2018} and the Ophiuchus DIsc Survey Employing ALMA (ODISEA) \citep{cieza2019,cieza2021} have shown that optically thick substructures are ubiquitous in bright and extended disks. This finding has sparked further interest in understanding the physical processes that give rise to such substructures and their implications for disk evolution and planet formation. Among the various mechanisms proposed to explain the origin of substructures in protoplanetary disks (see e.g., \citealt{baeppvii}), the most widely believed scenario is that substructures observed in Class II disks are created by the presence of planets or protoplanets within the disk. This hypothesis is supported by kinematic evidence \citep{Teague2018, Pinte2018, Izquierdo2022new} and by detection of planets within gaps, such as in PDS70 \citep{Muller2018, Keppler2018}. The interest in protoplanetary disk substructures extends beyond the observational findings. Indeed, it also arises from theoretical considerations, as they may help address several open key problems in disk evolution and planet formation theories. In particular, the fact that dust radially drifts inward in a disk as a result of the different velocities between the dust and gas in the disk \citep{weidenschilling1977aerodynamics} is a well-known open question of the field. Solid particles migrate too quickly in smooth disks due to the high efficiency of the radial drift \citep{takeuchi2002radial,takeuchi2005attenuation}, challenging the formation of planetesimals and leading to mismatches in spectral index predictions \citep{Birnstiel2010b,pinilla2013explaining} and size-luminosity distributions \citep{Tripathi2017,Andrews2018a,Rosotti2019} compared to observations. Substructures, particularly the presence of local maxima in the pressure profile of the disk, can mitigate solid particle migration, or completely trap them at the pressure maximum \citep{pinilla2012trapping,pinilla2013explaining}. If the substructure trapping mechanism is effective, solid particles accumulate at the pressure maximum enhancing the local dust-to-gas ratio \citep{drkazkowska2016close,yang2017concentrating}, thus creating a favorable region for planetesimal formation \citep{youdin2002planetesimal,youdin2007particle}. Additionally, substructures provide an ideal environment for planet formation by promoting processes like streaming instability and pebble accretion \citep{morbidelli2020,guilera2020,chambers2021,lau2022rapid,jiang2023efficient}. They also offer solutions to the migration problem \citep[e.g.,][]{matsumura2017n,liu2019growth,bitsch2019formation,lau2024can}, such as explaining the location of Jupiter within the Solar System \citep{lau2024sequential}.\\
\\
The quick disappearance of large-sized dust particles in disks is not the only challenging problem of disk evolution. The gas component is also evolving with time, and the gas size, commonly measured as the radius enclosing $68\,\%, 90\,\%$, or $95\,\%$ of the molecular emission of CO, is also studied as a way to probe disk evolution \citep{ansdell2018alma,trapman2020observed,trapman2022effect}. Different mechanisms of disk evolution predict different pathways for gas disk sizes. In a scenario where accretion is driven by viscosity, disk sizes should globally increase with time (the so-called 'viscous spreading' effect) to conserve angular momentum, while such expansion is not required in case of magnetohydrodynamical disk winds removing angular momentum from the disk \citep[for a review see e.g.,][]{Manara2023}. This is not the end of the story, as environmental effects such as multiplicity \citep[e.g.,][]{zagaria2023dust}, external and internal photoevaporation \citep[e.g.,][]{clarke2007photoevaporation,winter_photoev_2022,Anania_2025} can reduce or modify the size of a disk, with substantial consequences for the evolution of disks.

A way to test with the same models the efficiency of radial drift and the mechanism driving disk evolution is to focus on the evolution of the dust and gas sizes of protoplanetary disks at the same time, value studied observationally by several authors in different star-forming regions (\citealt{sanchis2021measuring} for Lupus, \citealt{long2022gas} for Taurus). For the measured sources, the vast majority of the disks in the sample show a ratio $R_{\mathrm{CO}}/R_{\mathrm{dust}}$ between 2-4, with a few (about 15$\%$) outliers with $R_{\mathrm{CO}}/R_{\mathrm{dust}} > 4$. This hints at a similar evolution mechanism for the two star-forming regions. However, theoretical analysis \citep{toci2021secular} modeling the evolution of protoplanetary disks assuming viscous evolution, grain growth and pure radial drift find that, for a broad range of disks values, the value of $R_{\mathrm{CO}}/R_{\mathrm{dust}}$ becomes > 5 after a short time.
Substructures have also been proposed as a possible solution to this discrepancy between the observed and simulated gas-to-dust size ratios in protoplanetary disks. \cite{toci2021secular} invoke the presence of unresolved or undetected substructures in most (or all) protoplanetary disks as the most likely explanation for reconciling the simulated gas-to-dust size ratios with the observed values. Indeed, the discrepancy between the small observed size ratios (i.e., $R_{\mathrm{CO}}/R_{\mathrm{dust}}<5$) and the high simulated size ratios (i.e., $R_{\mathrm{CO}}/R_{\mathrm{dust}}>5$) may potentially be addressed by increasing the disk dust size, $R_{\mathrm{dust}}$, through the presence of substructures within the disk.
Performing a population synthesis study, \cite{delussu2024population} have shown that substructures, and their rapid formation, are needed to produce small values of the spectral index in the range of the observed ones, and they allow reproducing the observed distributions for both spectral index and size-luminosity.

In this work, we aim to extend the population synthesis study of \cite{delussu2024population} by investigating whether, as proposed by \cite{toci2021secular}, substructures can provide a solution to the gas-to-dust size ratio problem. Moreover, we aim to understand whether it is possible to match the gas-dust size, size-luminosity, and spectral index distributions at the same time and what initial disk parameters are required for that.
Furthermore, we consider the impact of the stellar formation environment, exploring whether the external photoevaporation of the disk caused by the ultraviolet (UV) radiation emitted by OBA-type stars may explain the  $R_{\mathrm{CO}}/R_{\mathrm{dust}}$ observed in the population of disks in the Lupus region.
The UV radiation induces outside-in photoevaporation of the outermost disk regions, which are poorly gravitationally bound to the central star (e.g. \citeauthor{Adams_2004}\citeyear{Adams_2004}, \citeauthor{winter_photoev_2022}\citeyear{winter_photoev_2022}).
The far-ultraviolet (FUV) component dominates over the extreme ultraviolet (EUV) in the nearby regions in $\sim$200 pc from the Sun, which host few massive stars (mainly of late-type B and early-type A).
However, indirect evidence and models show that even a moderate FUV field (1-10 $\mathrm{G}_{0}$) can significantly affect disk evolution (e.g. \citeauthor{SODA_2023}\citeyear{SODA_2023}), with photoevaporative mass loss rates depending on the stellar mass and disk parameters (e.g. \citeauthor{FRIEDv2}\citeyear{FRIEDv2}).\\
\\
This paper is structured as follows: in Sect. \ref{methods}, we describe our computational models for the evolution of the disk and introduce the analysis method exploited to compare to disk observations. Section \ref{results} introduces the main results obtained and the comparison to the observed distributions. We first introduce the population synthesis results obtained with the two-pop-py model and then the results obtained for a small set of test disks using the code \texttt{DustPy} and in the presence of external photoevaporation. In Section \ref{sec:discussion} we discuss our results and their implications. Section \ref{conclusions} presents our conclusions.

\section{Methods}\label{methods}

The two-population model (two-pop-py) by \cite{birnstiel2012simple} and \cite{birnstiel2015dust} and \texttt{DustPy} code \citep{stammler2022dustpy} were employed to perform numerical simulations to describe the gas and dust evolution in a protoplanetary disk. In the following, we will describe the main characteristics of the two-pop-py \footnote{two-pop-py v1.1.4 was used for the simulations presented in this work} model and \texttt{DustPy} \footnote{\texttt{DustPy} v1.0.6 was used for the simulations presented in this work} code, and describe how external photoevaporation has been implemented in \texttt{DustPy}. The observables examined in this work will also be introduced, along with the process applied to evaluate each of them for every simulated disk.

\subsection{Two-pop-py model}\label{sec:twopoppy}

Two-pop-py is a tool that is well suited for disk population studies since it captures the dust surface density evolution, the viscous evolution of the gas and the particle size with good accuracy. Being based on a set of simple equations, it allows to perform a single simulation quickly (order of seconds), making it computationally efficient and allowing for the execution of large numbers of simulations within a reasonable amount of time. The two-pop-py model, as implemented and described in \cite{delussu2024population}, has been used in this work. For an in-depth discussion, we refer to \cite{delussu2024population}; a summary of its key elements is outlined as follows:
\begin{itemize}
    \item The protoplanetary gas disk is evolved according to the viscous disk evolution equation \citep{Lust1952,Lynden-Bell1974} using the turbulent effective viscosity as parameterized in \citet{shakura1973black}.
    \item Adopting the two-population model described in \cite{birnstiel2012simple} we evolve the dust surface density assuming that the small dust is tightly coupled to the gas while the large dust particles can decouple from it and drift inward.
    \item The initial gas surface density follows the \cite{lynden1974evolution} self-similar solution,\begin{equation}\label{eq:surface_density}
              \Sigma_{g}(r) = \Sigma_{0} \left( \frac{r}{\rc} \right)^{-p} \exp \left[ - \left( \frac{r}{\rc} \right)^{2-p} \right],
          \end{equation}
          where the normalization parameter $\Sigma_{0}=(2 - p) M_{disk}/2\pi \rc^{2}$ is set by the initial disk mass $M_{disk}$, $\rc$ denotes the so-called characteristic radius of the disk and $p$ the initial surface density power law exponent. $p$ has been set to $1$ for the initial profiles of all the disks.
    \item We have adopted a passive irradiated disk temperature model \citep{kenyon1996magnetic} with a disk floor temperature set to \SI{10}{K}. No viscous heating or other processes have been considered.
    \item Both, disks without (referred to as smooth disks) and with substructure have been examined. Substructure has been modeled as a gap due to the presence of a planet inserted in the disk. To mimic the presence of a planetary gap we have subdivided the $\alpha$-viscosity parameter into two different values: $\alpha_{\mathrm{gas}}$ and $\alpha_{\mathrm{dust}}$. The presence of the planet has been modeled as a local variation of the $\alpha_{\mathrm{gas}}$ parameter. Given the inverse proportionality between $\alpha_{\mathrm{gas}}$ and $\Sigma_{\mathrm{g}}$ in a steady-state regime, a bump in the $\alpha_{\mathrm{gas}}$ profile results in a gap in the $\Sigma_{\mathrm{g}}$ profile, effectively simulating the presence of a planetary gap. Moreover, this approach preserves the viscous evolution of $\Sigma_{\mathrm{g}}$. We have adopted the \cite{kanagawa2016mass} prescription to model the planetary gaps created in the disk.

\end{itemize}
In this work, the 1D disk has been spatially modeled with a logarithmic radial grid that ranges from \SI{0.05}{au} to \SI{10000}{au}. The main parameters of the grid model are presented in Table \ref{table:1}.
A total of $10^{5}$ simulations were performed for each population synthesis. To map the entire parameter space, the set of initial conditions adopted for each disk has been constructed randomly drawing each parameter from a probability distribution function (PDF). As in \cite{delussu2024population}, the main parameters that have been taken into account to describe both smooth and substructured disks are: disk mass ($M_{\rm{disk}}$), stellar mass ($M_{\rm{star}}$), disk characteristic radius ($\rc$), viscosity parameter ($\alpha$) and fragmentation velocity ($v_{\rm{frag}}$). The stellar mass values, have been drawn from a functional form of the IMF proposed by \cite{maschberger2013function}. Substructured disks have been characterized by three additional parameters: the mass of the planet creating the gap ($m_{\rm{p}}$), the time ($t_{\rm{p}}$) at which the planet was inserted, and the position ($r_{\rm{p}}$) of the planet within the disk. Table \ref{table:2} shows the range that has been adopted for each parameter and the corresponding PDFs adopted for each of them.
\begin{table}
    \tiny
    \caption{Fixed parameters adopted for two-pop-py model.}
    \label{table:1}
    \centering
    \begin{tabular}{c c c}
        \hline\hline
        Parameter                         & Description                 & Value or Range      \\
        \hline
        $\Sigma_{\rm{d}}/\Sigma_{\rm{g}}$ & initial dust-to-gas ratio   & 0.01                \\
        $\rho_{\rm{s}}\ [\rm{g/cm^{3}}]$  & particle bulk density       & 1.7 (Ricci opacity) \\
                                          & (no porosity)                                     \\
        $p$                               & initial surface density     & 1                   \\
                                          & power law exponent          &                     \\
        $r\ [\rm{au}]$                    & grid extent                 & 0.05-10000          \\
        $n_{\rm{r}}\ [\rm{cells}]$        & grid resolution             & 2000                \\
        $t\ [\rm{Myr}]$                   & duration of each simulation & 3                   \\
        \hline
    \end{tabular}
\end{table}

\begin{table}
    \tiny
    \caption{Disk initial parameters adopted for two-pop-py population synthesis.}
    \label{table:2}
    \centering
    \scalebox{0.95}{
        \begin{tabular}{c c c c}
            \hline\hline
            Parameter                             & Description            & Range             & PDF         \\
            \hline
            $\alpha$                              & viscosity parameter    & $10^{-4}-10^{-2}$ & log uniform \\
            $M_{\rm{disk}}\ [M_{\mathrm{star}}]$  & initial disk mass      & $10^{-3}-0.5$     & log uniform \\
            $M_{\mathrm{star}}\ [\rm{M_{\odot}}]$ & stellar mass           & $0.2-2.0$         & IMF         \\
            $\rc\ [\rm{au}]$                      & characteristic radius  & $10-230$          & log uniform \\
            $v_{\rm{frag}}\ [\rm{cm/s}]$          & fragmentation velocity & $200-2000$        & uniform     \\
            $m_{\rm{p}}\ [\rm{M_{\oplus}}]$       & planet mass            & $1-1050$          & uniform     \\
            $r_{\rm{p}}\ [\rc]$                   & planet position        & $0.05-1.5$        & uniform     \\
            $t_{\rm{p}}\ [\rm{Myr}]$              & planet formation time  & $0.1-0.4$         & uniform     \\
            \hline
        \end{tabular}}
    \\
    \tablefoot{Disk initial parameters and corresponding probability distribution function (PDF) from which their value is drawn for each single simulation. The drawing of the $m_{\rm{p}}$ value was performed after the $M_{\rm{disk}}$ to impose the further physically reasonable restriction of $m_{\rm{p}}<M_{\rm{disk}}$. In the case of multiple substructures, we applied the following constraint: $\sum_{i} m_{\rm{p,i}}<M_{\rm{disk}}$ for all planets $i$.
    }
\end{table}

\subsection{DustPy and external photoevaporation}
\begin{table}
    \tiny
    \caption{Disk initial parameters adopted for \texttt{DustPy} simulations.}
    \label{table:3}
    \centering
    \scalebox{0.95}{
        \begin{tabular}{c c c}
            \hline\hline
            Parameter                             & Description            & Value                    \\
            \hline
            $M_{\mathrm{star}}\ [\rm{M_{\odot}}]$ & stellar mass           & 0.3, 1                   \\
            $M_{\rm{disk}}\ [M_{\mathrm{star}}]$  & initial disk mass      & $0.01$, $0.1$            \\
            $\rc\ [au]$                           & characteristic radius  & 20, 50, 100              \\
            $\alpha$                              & viscosity parameter    & $10^{-3.5}$, $10^{-2.5}$ \\
            $v_{\rm{frag}}\ [cm/s]$               & fragmentation velocity & 1000                     \\
            $m_{\rm{p}}\ [\rm{M_{J}}]$            & planet mass            & 1                        \\
            $r_{\rm{p}}\ [\rc]$                   & planet position        & 0.5, 0.7, 1, 1.2         \\
            $F_{\mathrm{FUV}}$ [$\mathrm{G}_{0}$] & FUV flux               & 4                        \\
            \hline
        \end{tabular}}
    \\
    \tablefoot{Disk initial parameters used for \texttt{DustPy} simulations of disk subject to viscous evolution and external photoevaporation. The FUV flux corresponds to the average in the Lupus region and is constant during the simulation.
    }
\end{table}
Given that external photoevaporation is not yet available in two-pop-py, to perform numerical simulations of disks undergoing external photoevaporation, we used the existing and tested implementation of \texttt{DustPy} \citep{stammler2022dustpy} integrated with an external module that includes the effect of an external FUV field \citep{Garate_2024,Anania_2025_AGEPROVIII}.
\texttt{DustPy} simulates the radial evolution of gas and dust in a viscous disk, considering a distribution of particle species subject to advection, diffusion, grain growth by coagulation and fragmentation based on \citet{Birnstiel2010}. $\Sigma_{\mathrm{g}}$, evolves accordingly with the \cite{lin1986tidal} and \cite{trilling1998orbital} advection-diffusion equation under the influence of axis-symmetrical torque deposition, where we included an extra term to account for the loss of material in external winds:
\begin{equation}
    \frac{\partial \Sigma_\mathrm{g}}{\partial t}=\frac{3}{r}\frac{\partial}{\partial r}\left[ r^{1/2} \frac{\partial}{\partial r}(\nu \Sigma_\mathrm{g} r^{1/2}) -\frac{2\Lambda\Sigma_\mathrm{g}}{3\Omega_\mathrm{K}} \right] - \dot{\Sigma}_{\mathrm{ext}},
    \label{eq:diffusion_eq}
\end{equation}
where $\nu=\alpha c_\mathrm{s}^2/\Omega_\mathrm{K}$ is the viscosity, $\Lambda$ is the specific angular momentum injection rate, and $\dot{\Sigma}_{\mathrm{ext}}$ accounts for the rate of mass lost in external photoevaporative winds.
The photoevaporative mass loss rate is evaluated performing a bi-linear interpolation of the FRIEDv2 grid \citep{FRIEDv2}, where we assumed ISM-like Polycyclic Aromatic Hydrocarbons (PAHs) for ISM-like dust. Specifically, following the numerical implementation of \citet{sellek_fried}, the grid is interpolated fixing the stellar mass and the FUV flux at the outer disk edge, and using the disk surface density at each radial position. At each time step, the disk radial position corresponding to the maximum mass loss rate defines a truncation radius (which corresponds to the optically thin/thick wind transition in the disk). To simulate the outside-in disk depletion caused by external photoevaporation, the disk material outside the truncation radius is removed accordingly to \citet{sellek_fried}.
The model takes into account that external photoevaporation influences the dust component, where dust grains smaller than a certain size limit, which depends on the stellar and disk parameters, are entrained in winds \citep{facchini_2016}.
The study of \citet{Garate_2023} showed that dust substructures that may form in disks survive the outside-in disk depletion only if they are placed inside the truncation radius of external photoevaporation.\\
\\
Due to the higher computational costs of \texttt{DustPy} compared to two-pop-py, we switched from a population synthesis study to a study of a test population of disks based on the constraints on the parameter space placed by two-pop-py simulations \citep{delussu2024population} of not-photoevaporating disks for the initial conditions needed to reproduce the observed distributions of spectral index and size-luminosity. The extent of the radial grid, the dust particle distribution and the fragmentation velocity are set as in two-pop-py simulations.
In total, we explored the parameter space by performing 27 simulations, considering all possible combinations of the parameter values listed in Table \ref{table:3}.
In the simulations including external photoevaporation, we assumed that the disks are subject to a constant external FUV flux of \SI{4}{\mathrm{G}_{0}}, which is the average FUV flux experienced by disk-hosting stars in the Lupus star-forming region \citep{Anania_2025}.
The simulated disks are let evolve up to the age of \SI{3}{Myr} to compare with the Lupus disk population.\\
\\
As for the study conducted with two-pop-py (Sect. \ref{sec:twopoppy}), the substructures in \texttt{DustPy} were also modeled as a gap arising from the presence of a planet inserted in the disk. We imposed the planetary gap through the radial gas flux induced by a torque profile, rather than modifying the viscosity $\alpha$-parameter. This approach avoids the issue of unrealistically high gas velocities with a deep planetary gap, which could arise from low gas surface density caused by disk dissipation due to external photoevaporation. Appendix~\ref{appendix:dustpy_vs_twopoppy} shows a comparison between \texttt{DustPy} and two-pop-py for substructured test disks. While minor differences are observed near the planet-induced substructure and modestly affect the spectral indices, the dust and gas disk radii, the main observables considered in this study, are in good agreement, and the main conclusions are therefore preserved. Given its more detailed treatment of substructure formation, we will adopt \texttt{DustPy} in the following.\\
The angular momentum injection rate $\Lambda$ in Eq. (\ref{eq:diffusion_eq}) is derived by assuming a steady state with the target surface density profile and an unchanged steady-state accretion rate. Neglecting the photoevaporation term, Eq. (\ref{eq:diffusion_eq}) can be rewritten as
\begin{equation}
    \frac{\partial \Sigma_\mathrm{g}}{\partial t}=\frac{3}{r}\frac{\partial}{\partial r}\left[ r \Sigma_\mathrm{g}( v_\nu+v_\Lambda) \right],
\end{equation}
with the accretion velocity
\begin{equation}\label{eq:visc_vel}
    v_\nu=-\frac{3\nu}{r}\frac{\partial\log}{\partial \log r} \left( \Sigma_\mathrm{g}r^{1/2}\nu \right),
\end{equation}
and the additional velocity due to the torque injection
\begin{equation}
    v_\Lambda=\frac{2\Lambda}{\Omega_\mathrm{K}r}.
\end{equation}
The condition for the planetary gap to not impede disk accretion can be expressed as
\begin{equation}\label{eq:steady}
    \Sigma_\mathrm{g,eq} (v_{\nu,\mathrm{eq}}+v_\Lambda)=\Sigma_\mathrm{g,0}v_{\nu,0},
\end{equation}
where $\Sigma_\mathrm{g,0}$ represents the unperturbed gas surface density, $v_{\rm{\nu,0}}$ is the unperturbed accretion velocity, $\Sigma_\mathrm{g,eq}$ and $v_{\nu,\mathrm{eq}}$ denote the gas surface density and accretion velocity at the steady state with the imposed gap. Starting from Eq. (\ref{eq:visc_vel}), defining the target surface density profile as
\begin{equation}
    f(r) \equiv \frac{\Sigma_\mathrm{g,eq}}{\Sigma_\mathrm{g,0}},
\end{equation}
and given that $\partial(\Sigma_\mathrm{g,0}\nu)/\partial r=0$ under the assumption of a constant disk accretion rate, $v_{\nu,\mathrm{eq}}$ is given by
\begin{align}
    v_{\nu,\mathrm{eq}} & =-\frac{3\nu}{r}\frac{\partial\log}{\partial \log r} \left( f\Sigma_\mathrm{g,0}r^{1/2}\nu \right) \\
                        & =-\frac{3\nu}{r}\left(\frac{\partial\log f}{\partial \log r} +\frac{1}{2}\right),
\end{align}
and $v_{\nu,0}$ is
\begin{equation}
    v_{\nu,0}=-\frac{3\nu}{2r}.
\end{equation}
Exploiting Eq. (\ref{eq:steady}), it follows that
\begin{align}
    v_\Lambda= & \frac{v_{\nu,0}}{f(r)}-v_{\nu,\mathrm{eq}}                                                     \\
    =          & -\frac{3\nu}{r}\left(\frac{1}{2f}-\frac{\partial\log f}{\partial \log r} - \frac{1}{2}\right).
\end{align}
To implement this in \texttt{DustPy}, an external source is applied and calculated with $v_\Lambda$, following the procedure described in \cite{lau2025formation}. For the profile $f$, we adopted the planetary gap profile by \cite{kanagawa2016mass}, as for the two-pop-py model (Sect. \ref{sec:twopoppy}).

\subsection{Observables}

Post-processing the resulting profiles (surface density, maximum grain size, and disk temperature) of the simulated disks, we obtained the simulated population distribution for the disk dust radii, disk gas radii, millimeter fluxes, and spectral indices. In the following we describe the process applied to evaluate each of these observables for every simulated disk.

\subsubsection{Millimeter fluxes}

Disk fluxes have been evaluated following the procedure described in \cite{delussu2024population}, that is, adopting \cite{miyake1993effects} scattering solution of the radiative transfer equation and the modified Eddington-Barbier approximation \citep{birnstiel2018}. \cite{delussu2024population} studied the effects of different opacities (Ricci compact \citealt{Rosotti2019}, DSHARP \citealt{birnstiel2018}, DIANA \citealt{woitke2016consistent}), and showed that only the Ricci compact opacity results in a match of the observed spectral index distribution. We therefore have adopted the Ricci compact opacity model for the analysis conducted in this paper.

\subsubsection{Dust radii}

One of the challenges in characterizing protoplanetary disks is defining their size \citep[see][as a recent review]{miotello2023setting}. As discussed in \cite{Tripathi2017} and \cite{rosotti2019time}, the characteristic radius $\rc$ is not an optimal choice as a size indicator for disks. In large sample surveys the millimeter surface brightness profile of observed disks is measured, thus, we have followed the procedure of defining an empirical radius, commonly referred to as the effective disk radius $R_{\rm eff}$, which represents the radius encompassing a specified fraction of the total flux emitted by the disk. The two most commonly adopted choices are $R_{68\%}$ and $R_{90\%}$, the radii enclosing $68\%$ and $90\%$ of the total disk flux, respectively. As shown in \cite{sanchis2020demographics}, $R_{68\%}$ is a safer choice, as it is usually characterized by a smaller dispersion. Based on the radii provided in the observed samples; $R_{68\%}$ has been evaluated for the comparison to the observed size-luminosity distribution \citep{Andrews2018a}, while $R_{90\%}$ has been evaluated for the comparison to the observed gas and dust radii distribution \citep{sanchis2021measuring}.

\subsubsection{Gas radii}

Estimating disk gas radii is even more challenging than estimating dust radii, as molecular emission lines are faint in the outer regions of protoplanetary disks. Additionally, deep gas observations are rarer as they are more time-consuming than dust continuum observations. Given their relatively brighter and optically thick emission, CO rotational emission lines, particularly $\ce{^{12}CO}$ lines, are the most commonly used tracers for disk gas radii estimates. Given that we compared our simulated disks to the observed size distribution of disks of the Lupus star-forming region presented in \cite{sanchis2021measuring}, which provides the radius enclosing the $90\%$ of the total flux emission of the $\ce{^{12}CO}$ tracer, we assumed $\ce{^{12}CO}$ as disk gas size tracer and $R_{\mathrm{CO},90\%}$ as the gas radius for each of the simulated disks. As shown in \cite{toci2023analytical}, $R_{\mathrm{CO},90\%}$ is well-approximated by the radial location where the disk surface density equals a certain critical value, inhibiting $\ce{^{12}CO}$ self-shielding.\\
\\
We evaluate the disk gas radius $R_{\mathrm{CO},90\%}$ starting from the gas critical column density threshold formula obtained by \cite{trapman2023large}:
\begin{equation}\label{trapman_ngascrit}
    \mathrm{N_{gas,crit}} = 10^{21.27-0.53log_{10}\mathrm{L_{\star}}}\left(\frac{\mathrm{M_{gas}}}{\mathrm{M_{\odot}}}\right)^{0.3-0.08log_{10}\mathrm{L_{\star}}} \mathrm{cm^{-2}}.
\end{equation}
Given the gas critical column density threshold, the critical surface density of the $\ce{^{12}CO}$ tracer is given by
\begin{equation}\label{sig_critgas}
    \Sigma_{\mathrm{crit}} = 10\mu_{\mathrm{gas}} N_{\rm{gas,crit}}
\end{equation}
where $\mu_{\mathrm{gas}}$ is the mean molecular weight of the gas in the disk; we adopted the standard value for protoplanetary disks (i.e., $\mu_{\mathrm{gas}} \sim 2.3 \mu_{\mathrm{H}} = 1.15 \mu_{\mathrm{H_{2}}}$). As in Eq. (\ref{trapman_ngascrit}) the initial carbon abundance is assumed to be $10^{-4}$; we account for extra carbon depletion due to CO freeze-out and photodissociation by multiplying the critical surface density by a factor \SI{10}{}. The gas radius $R_{\mathrm{CO,90\%}}$ was then evaluated as the outermost radius for which the gas surface density profile obtained for each simulated disk equals its critical surface density.

\subsubsection{Spectral index}

The spectral index is defined as the slope of the (sub-)mm SED of the dust emission
\begin{equation}\label{spectral_def}
    \alpha_{mm} = \frac{dlogF_{\nu}}{dlog \nu},
\end{equation}
where $F_{\nu}$ is the disk-integrated flux at a given frequency $\nu$\footnote{All fluxes will be given in frequency space ($F_{\nu}$, typically in units of Jy), even though the subscript is stating the corresponding wavelength instead of the frequency.}, thus $\alpha_{mm}$ is the disk-integrated spectral index. Since we typically work with frequencies that are very close to each other, Eq. (\ref{spectral_def}) can be written as
\begin{equation}\label{spectral_eq}
    \alpha_{mm} = \frac{log(F_{\nu,2}/F_{\nu,1})}{log(\nu_{2}/\nu_{1})}.
\end{equation}
Eq. (\ref{spectral_eq}) has been applied to determine the spectral index for every simulated disk. Since we compare our simulated spectral index distribution to the observed Lupus region sample adopted in \cite{tazzari2021first}, which is a collection of disks detected at 0.89mm \citep{ansdell2016alma} and 3.1mm \citep{tazzari2021first}, we have considered $\lambda_{2}=\SI{0.89}{mm}$ and $\lambda_{1}=\SI{3.10}{mm}$.

\begin{figure*}[!htb]
    \centering
    \includegraphics[scale=0.45]{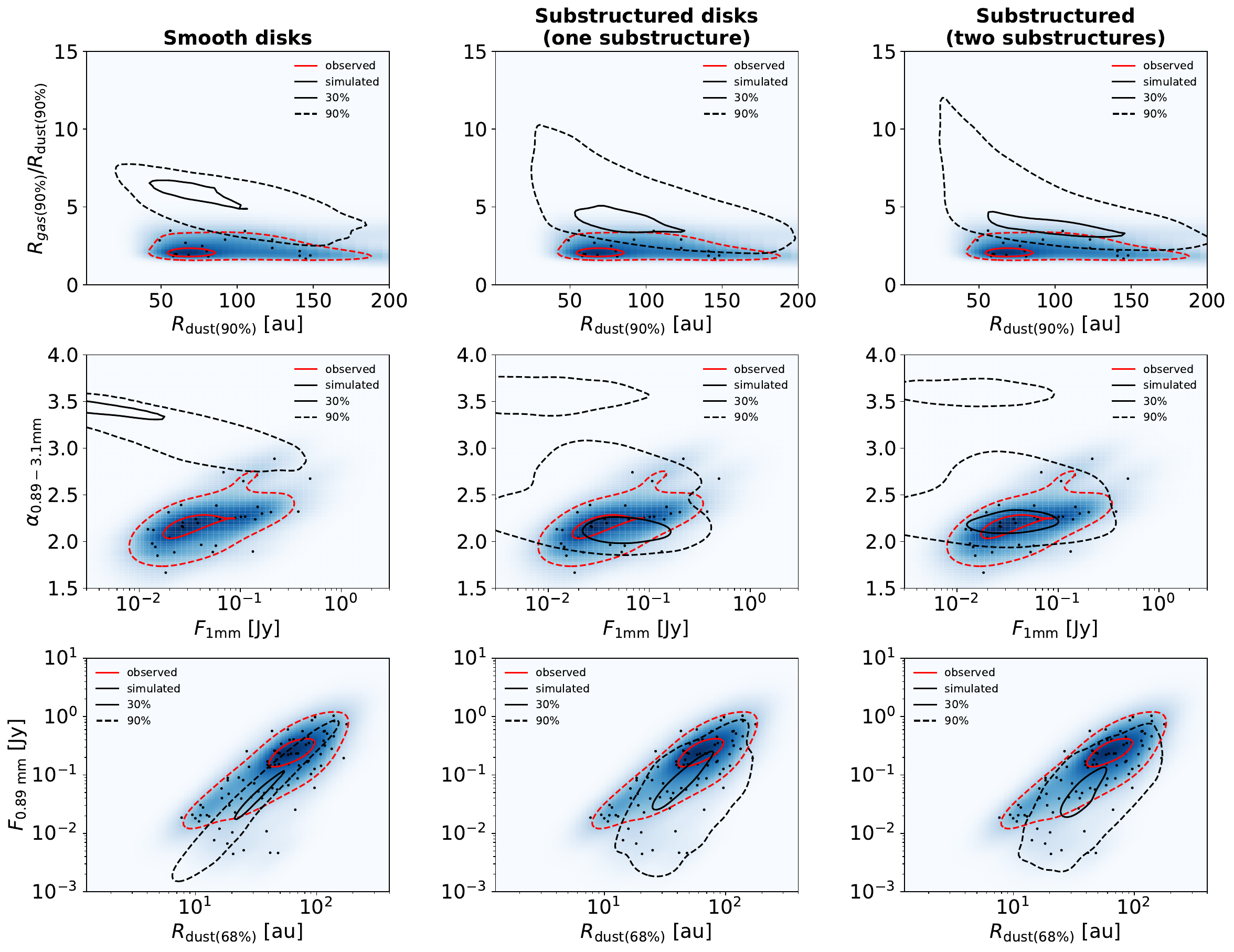}
    \caption{Gas-dust size distribution (first row), spectral index distribution (second row) and size-luminosity distribution (third row) for the parameter space of the initial conditions selecting disks with a spectral index $0 \leq \alpha_{\mathrm{0.89-3.1mm}}\leq 4$, $\SI{e-3}{Jy}\leq F_{\rm{1mm}}\leq \SI{10}{Jy}$, $\SI{e-3}{Jy}\leq F_{\rm{0.89mm}}\leq \SI{10}{Jy}$, $10^{0.1}\,\si{au}\leq R_{\rm{dust(68\%)}}\leq 10^{2.6}\,\si{au}$, $\SI{1}{au}\leq R_{\rm{dust(90\%)}}\leq \SI{200}{au}$ and $0.1\leq R_{\rm{gas(90\%)}}/R_{\rm{dust(90\%)}}\leq 20$.\\
    Left plots: smooth disks. Middle plots: substructured disks with one planet randomly inserted in a range between 0.1-0.4 Myr from the start of the disk evolution. Right plots: substructured disks with two planets randomly inserted in a range between 0.1-0.4 Myr (innermost planet) and between 0.5-0.8 Myr (outermost planet) from the start of the disk evolution.\\
    Heatmap of the observed disks with the black dots representing each single observed disk. The black and red lines refer to the simulated results and the observational results respectively. In particular, the continuous lines encompass $30\%$ of the cumulative sum of the disks produced from the simulations or observed. The dashed lines encompass instead the $90\%$.}
    \label{fig:smooth_vs_substr}
\end{figure*}

\section{Results}\label{results}

The following section contains the main results obtained through our analysis. Subsection \ref{twopoppy_res} presents the results obtained for the population synthesis produced with two-pop-py for the reproducibility of the gas and dust size observed distribution. In Subsection \ref{dustpy_res} we show the results obtained for a test population of disks exploiting \texttt{DustPy} code with the implementation of external photoevaporation.
To compare the simulated distributions with the observed ones, a potential age spread of the simulated disks was considered. Specifically, for each simulated disk, the observables used to construct the overall simulated distributions were randomly selected from the snapshots at \SI{1}{Myr}, \SI{2}{Myr}, and \SI{3}{Myr}.
Our simulated gas and dust size distribution is compared to the observed sample reported in \cite{long2022gas} for the Lupus region, while the simulated spectral index distributions are compared to the observed sample from \cite{tazzari2021first}. The latter includes disks detected at 0.89mm \citep{ansdell2016alma} and 3.1mm \citep{tazzari2021first}. The simulated size-luminosity distribution is compared to the observed sample of the Lupus region reported in \cite{Andrews2018a}.

\subsection{Population synthesis results with two-pop-py}\label{twopoppy_res}

The main focus of the study, conducted using two-pop-py model, was to investigate the population synthesis results for the gas and dust size distribution of smooth and substructured disks, to assess the validity of the solution proposed by \cite{toci2021secular}, namely, that substructures can solve the discrepancy between models and observations. \autoref{fig:smooth_vs_substr} shows the clear difference that has been found between smooth and substructured disks. If we consider the entire parameter space of the initial conditions adopted and reported in Table \ref{table:2}, we observe the first important indications obtained through our simulations:
\begin{itemize}
    \item Smooth disks produce a large $R_{\mathrm{gas(90\%)}}/R_{\mathrm{dust(90\%)}}$ compared to observations. This result confirms and extends to the broader level of a disk population synthesis the result previously obtained by \cite{toci2021secular}.
    \item Substructured disks, either with one or two substructures, produce a size distribution shifted towards lower $R_{\mathrm{gas(90\%)}}/R_{\mathrm{dust(90\%)}}$. For both one and two substructures, the bulk of the population lies around $R_{\mathrm{gas(90\%)}}/R_{\mathrm{dust(90\%)}} \sim 4$.
    \item The presence of substructure(s) in the disks helps narrow the gap between theoretical models and observations, but it does not fully resolve the discrepancy.
\end{itemize}
For completeness, \autoref{fig:smooth_vs_substr} also presents the spectral index and size-luminosity distributions associated with the three types of disks examined. As already shown in \cite{delussu2024population}, only substructured disks are capable of producing spectral indices that populate the observed spectral index region. Therefore,  \autoref{fig:smooth_vs_substr} illustrates how smooth disks fail to reproduce two of the three observed distributions examined. However, it also highlights a tension for substructured disks between the favorable results for the spectral index and size-luminosity distributions, which can be simultaneously reproduced by restricting the parameter space of the initial conditions of the simulated disks \citep{delussu2024population}, and the discrepancy between simulated and observed size distribution.

\begin{figure*}[!htb]
    \centering
    \includegraphics[scale=0.38]{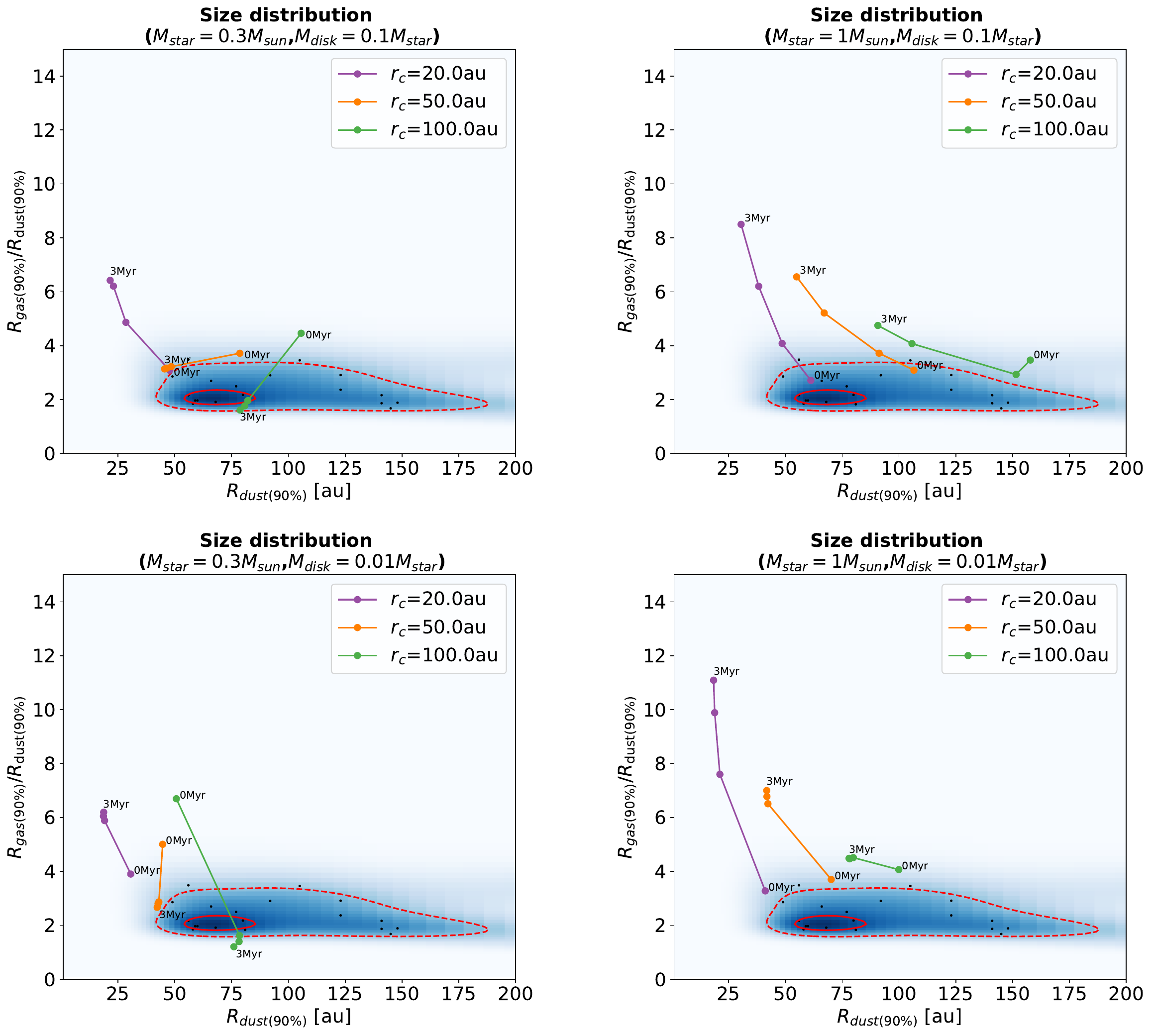}
    \caption{Evolution in the $R_{\rm{gas(90\%)}}/R_{\rm{dust(90\%)}}$ vs $R_{\rm{dust(90\%)}}$ space for some test substructured disks evolved with \texttt{DustPy} code with external photoevaporation ($F_{\rm{FUV}} = \SI{4}{\mathrm{G}_{0}}$) in a low viscosity regime ($\alpha=10^{-3.5}$) for different values of the characteristic radius $r_{\rm{c}}$. The points associated with each trajectory represent the snapshots taken at \SI{0}{Myr}, \SI{1}{Myr}, \SI{2}{Myr}, and \SI{3}{Myr}, respectively.}
    \label{fig:dustpy_lowviscosity}
\end{figure*}

\subsection{DustPy and external photoevaporation results}\label{dustpy_res}

Having determined that the mere presence of substructures in protoplanetary disks is insufficient to solve the size distribution problem, a further study was conducted to investigate whether external photoevaporation, in conjunction with substructures, may be the necessary ingredient to reconcile observations and simulations. Nevertheless, given that external photoevaporation is not yet available in two-pop-py, we switched from a population synthesis study to a study of a test population of disks conducted with \texttt{DustPy} where the effect of external photoevaporation is included in addition to viscous evolution. We assumed that the disks are exposed to a constant external FUV flux of 4 $\mathrm{G}_{0}$, which corresponds to the average FUV flux experienced by disk-hosting stars in the Lupus star-forming region \citep{Anania_2025}.
\autoref{fig:dustpy_lowviscosity} shows the behavior of some substructured test disks in the $R_{\mathrm{gas(90\%)}}/R_{\mathrm{dust(90\%)}}$ vs $R_{\mathrm{dust(90\%)}}$ space in the regime of low viscosity (i.e., $\alpha=10^{-3.5}$):
\begin{itemize}
    \item Right panels of Fig. \ref{fig:dustpy_lowviscosity} show that disks hosted by a solar mass star evolve towards high ratios of $R_{\mathrm{gas(90\%)}}/R_{\mathrm{dust(90\%)}}$. The disk decreases in $R_{\mathrm{dust(90\%)}}$ rapidly settling to a fixed value corresponding to the location of the substructure; nevertheless, $R_{\mathrm{gas(90\%)}}/R_{\mathrm{dust(90\%)}}$ increases over time because these disks experience viscous spreading or just slightly reduces their $R_{gas(90\%)}$ due to a small effect of the external photoevaporation.
    \item Left panels of Fig. \ref{fig:dustpy_lowviscosity} show that disks hosted by a star with $M_{\rm{star}}=\SI{0.3}{\mathrm{M_{\odot}}}$ and with a small characteristic radius $r_{\rm{c}}=\SI{20}{au}$ evolve towards high ratios of $R_{\mathrm{gas(90\%)}}/R_{\mathrm{dust(90\%)}}$ as they experience viscous spreading because their external photoevaporation truncation radius is placed too farther away ($\sim \SI{100}{au}$) for impacting on disk evolution.
    \item Left panels of Fig. \ref{fig:dustpy_lowviscosity} show that medium ($r_{\rm{c}}=\SI{50}{au}$) and large ($r_{\rm{c}}=\SI{100}{au}$) disks hosted by a star with $M_{\rm{star}}=\SI{0.3}{\mathrm{M_{\odot}}}$ evolve towards small ratios of $R_{\mathrm{gas(90\%)}}/R_{\mathrm{dust(90\%)}}$, falling close or inside the observed region.
\end{itemize}

\begin{figure*}[!htb]
    \centering
    \includegraphics[scale=0.4]{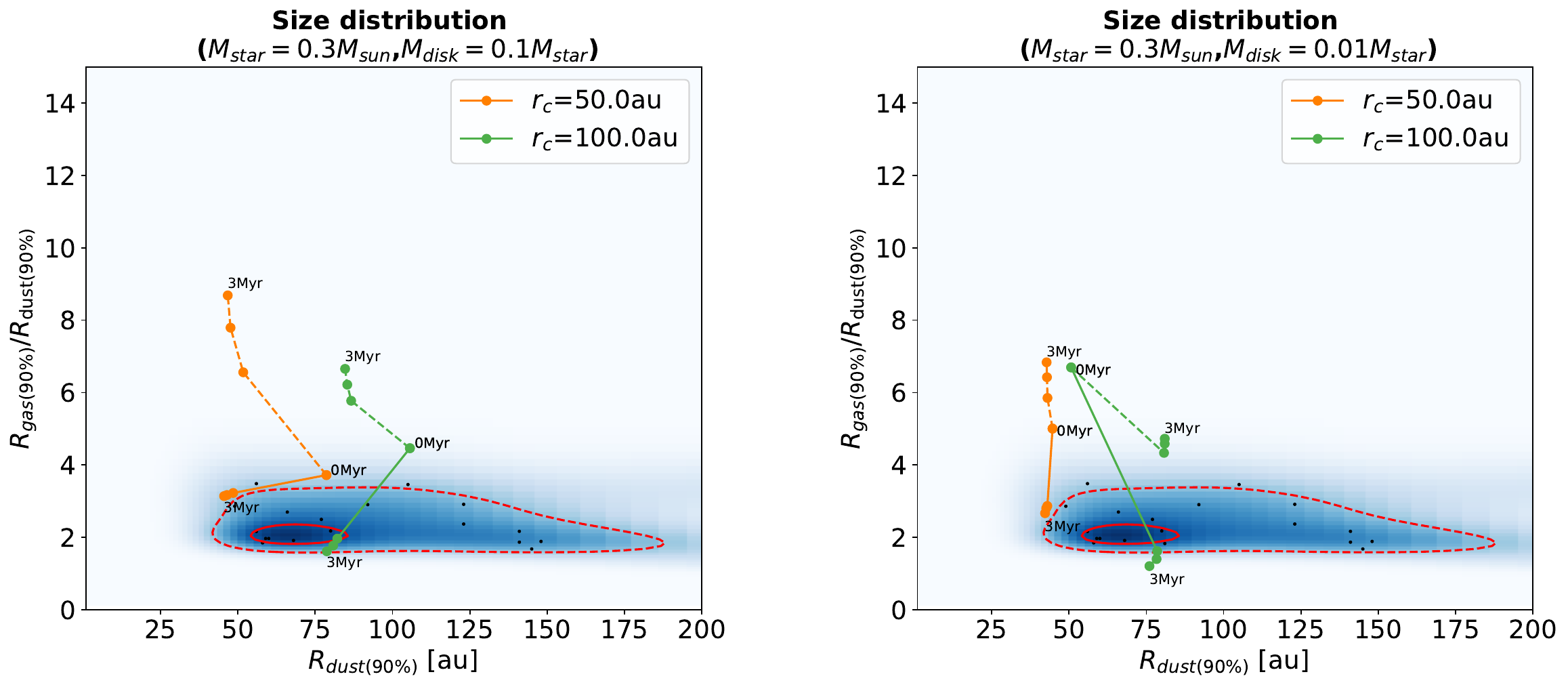}
    \caption{Evolution in the $R_{\mathrm{gas(90\%)}}/R_{\mathrm{dust(90\%)}}$ vs $R_{\mathrm{dust(90\%)}}$ space for some test substructured disks evolved with \texttt{DustPy} code in a low viscosity regime ($\alpha=10^{-3.5}$) for different values of the characteristic radius $\rc$. The points associated with each trajectory represent the snapshots taken at \SI{0}{Myr}, \SI{1}{Myr}, \SI{2}{Myr}, and \SI{3}{Myr}, respectively. Solid line: external photoevaporation ($F_{\rm{FUV}} = \SI{4}{\mathrm{G}_{0}}$). Dashed line: no external photoevaporation ($F_{\rm{FUV}} = \SI{0}{\mathrm{G}_{0}}$).}
    \label{fig:photo_vs_nophoto}
\end{figure*}

Having established that small-size disks (i.e., $r_{\rm{c}}=\SI{20}{au}$) and disks hosted by a solar-mass star do not exhibit the desired behavior, we focused on medium (i.e., $r_{\rm{c}}=\SI{50}{au}$) and large (i.e., $r_{\rm{c}}=\SI{100}{au}$) disk size cases hosted by a star of mass $M_{\rm{star}}=\SI{0.3}{\mathrm{M_{\odot}}}$.
\autoref{fig:photo_vs_nophoto} shows the impact of the external photoevaporation on the behavior of these disks. Disks completely revert their behavior in the absence of external photoevaporation moving towards high $R_{\mathrm{gas(90\%)}}/R_{\mathrm{dust(90\%)}}$ ratios as they experience viscous spreading in $R_{\mathrm{gas(90\%)}}$ in the absence of external photoevaporation. The first decrease in $R_{\mathrm{gas(90\%)}}/R_{\mathrm{dust(90\%)}}$ experienced by the disk with mass $M_{\rm{disk}}= \SI{0.01}{\mathrm{M_{star}}}$ and $r_{\rm{c}}=\SI{100}{au}$ is simply because $R_{\mathrm{dust(90\%)}}$ settles towards the location of the substructure which in this case happens to produce an increase in $R_{\mathrm{dust(90\%)}}$. However, also this disk is affected by spreading in $R_{\mathrm{gas(90\%)}}$, which combined with the stable $R_{\mathrm{dust(90\%)}}$ produced by the presence of the substructure leads to an increase of $R_{\mathrm{gas(90\%)}}/R_{\mathrm{dust(90\%)}}$ over time.  \autoref{fig:1substr_50rc} and \autoref{fig:1substr_100rc} show an extension and in-depth investigation of our discussion focused on the disk cases that have proven so far successful in their behavior in the $R_{\mathrm{gas(90\%)}}/R_{\mathrm{dust(90\%)}}$ vs $R_{\mathrm{dust(90\%)}}$ space.

\begin{itemize}
    \item Figure \ref{fig:1substr_50rc} shows that medium disks (i.e., $\rc = \SI{50}{au}$) experience a continuous reduction of their $R_{\mathrm{gas(90\%)}}/R_{\mathrm{dust(90\%)}}$ for all the substructures locations examined. However, if we extend our investigation to the behavior of these disks in the spectral index and size-luminosity space we immediately notice the emergence of some problems. All  'lighter' disks (i.e., with an $M_{\rm{disk}}= \SI{0.01}{\mathrm{M_{star}}}$) produce large values of spectral indexes and extremely low fluxes. The same problem is displayed by the more massive disks (i.e., with an $M_{\rm{disk}}=\SI{0.1}{\mathrm{M_{star}}}$) if the substructure is placed at or beyond the characteristic radius. Nevertheless, placing the substructure too far inside does not yield beneficial results, as demonstrated by the case $r_{\rm{p}} = \SI{0.5}{r_c}$. This configuration produces a smaller $R_{\mathrm{dust(90\%)}}$, which fails to sufficiently decrease the $R_{\mathrm{gas(90\%)}}/R_{\mathrm{dust(90\%)}}$ ratio.
    \item  Figure \ref{fig:1substr_50rc} shows that if the substructure is placed at an intermediate position (i.e., $r_{\rm{p}} = \SI{0.7}{r_c}$) and $M_{\rm{disk}}=\SI{0.1}{\mathrm{M_{star}}}$, the disk behaves well in all the parameter space taken into account. Nevertheless, it introduces a warning that we face a fine-tuning problem.
    \item Figure \ref{fig:1substr_100rc} shows that large disks (i.e., $\rc = \SI{100}{au}$) experience a continuous reduction of their $R_{\mathrm{gas(90\%)}}/R_{\mathrm{dust(90\%)}}$ only in the case in which the substructure is placed at $r_{\rm{p}} =\SI{ 0.5 }{r_c}$. Indeed, placing the substructure farther away causes it to be gradually removed by the external photoevaporation mechanism as it gets closer to the truncation radius location ($\sim \SI{110}{au}$). Furthermore, the erosion of the substructure due to external photoevaporation translates into severe discrepancies with respect to the observed spectral indices and fluxes as these disks start to behave as smooth disks. Despite its good behavior in the size and size-luminosity space, the $r_{\rm{p}} =\SI{ 0.5 }{r_c}$ fails to produce a spectral index in the observed range.
\end{itemize}

\section{Discussion}\label{sec:discussion}

\subsection{A tension between dust and gas radii}

The population study conducted by \cite{delussu2024population} revealed that substructured disks can simultaneously reproduce the observed spectral index and size-luminosity distributions by imposing simple constraints on the parameter space of their initial conditions. However, both the population study performed with two-pop-py (Sec. \ref{twopoppy_res}) and the study on a test population conducted with \texttt{DustPy} in the presence of external photoevaporation (Sec. \ref{dustpy_res}) exposed the difficulty in reproducing the observed ratios of $R_{\mathrm{gas(90\%)}}/R_{\mathrm{dust(90\%)}}$.

The results obtained from both studies presented in the previous sections demonstrate that both dust sizes and spectral indices can be reproduced if substructures are included. However, in those cases the gas radii are significantly overpredicted. While external photoevaporation can help to shrink the gas sizes, it in turn prevents the larger dust sizes to be realized by the models. In addition, external photoevaporation can even dissipate pressure traps that are near the truncation radius. The results indicate that viscosity works towards increasing the gas size while radial drift decreases the dust size. Substructures prevent the latter, but a mechanism is required that decreases the measured gas size without affecting the dust trapping. Conversely, when gas radii are reproduced, as in \cite{toci2021secular} by considering smooth disks, the dust radii are underestimated.

It was shown that combinations of initial conditions can be found where disks undergoing viscous evolution with external photoevaporation, yield results consistent with the observed ones. However, the limited range of initial conditions required to achieve such outcomes make this a fine-tuning problem.

If measurements of gas radii indeed trace the outer edges of the gas density while the dust substructure is almost always within a factor of 2 of this radius, it would suggest either that the outermost trap is somehow associated with the disk truncation radius, or that traps are so frequent within each disk that there always exists a trap near the gas outer radius.\\

In this work, we adopted a constant FUV flux of \SI{4}{\mathrm{G}_{0}}, neglecting the impact of dust extinction and cluster dynamics. Dust can efficiently shield disks from incoming UV radiation in the first 0.5-1 Myr of disk lifetime \citep[see e.g.,][]{Ali_extinction_2019,Qiao_extinction_2022}. Moreover, the FUV flux experienced by disks is not constant in time due to the relative motion of OBA stars and disks. However, since Lupus star-forming region does not contain early-type B and O-type stars, we can suppose that this last effect is not significantly compromising our results.

\begin{figure*}[tbh]
    \centering
    \includegraphics[scale=0.4]{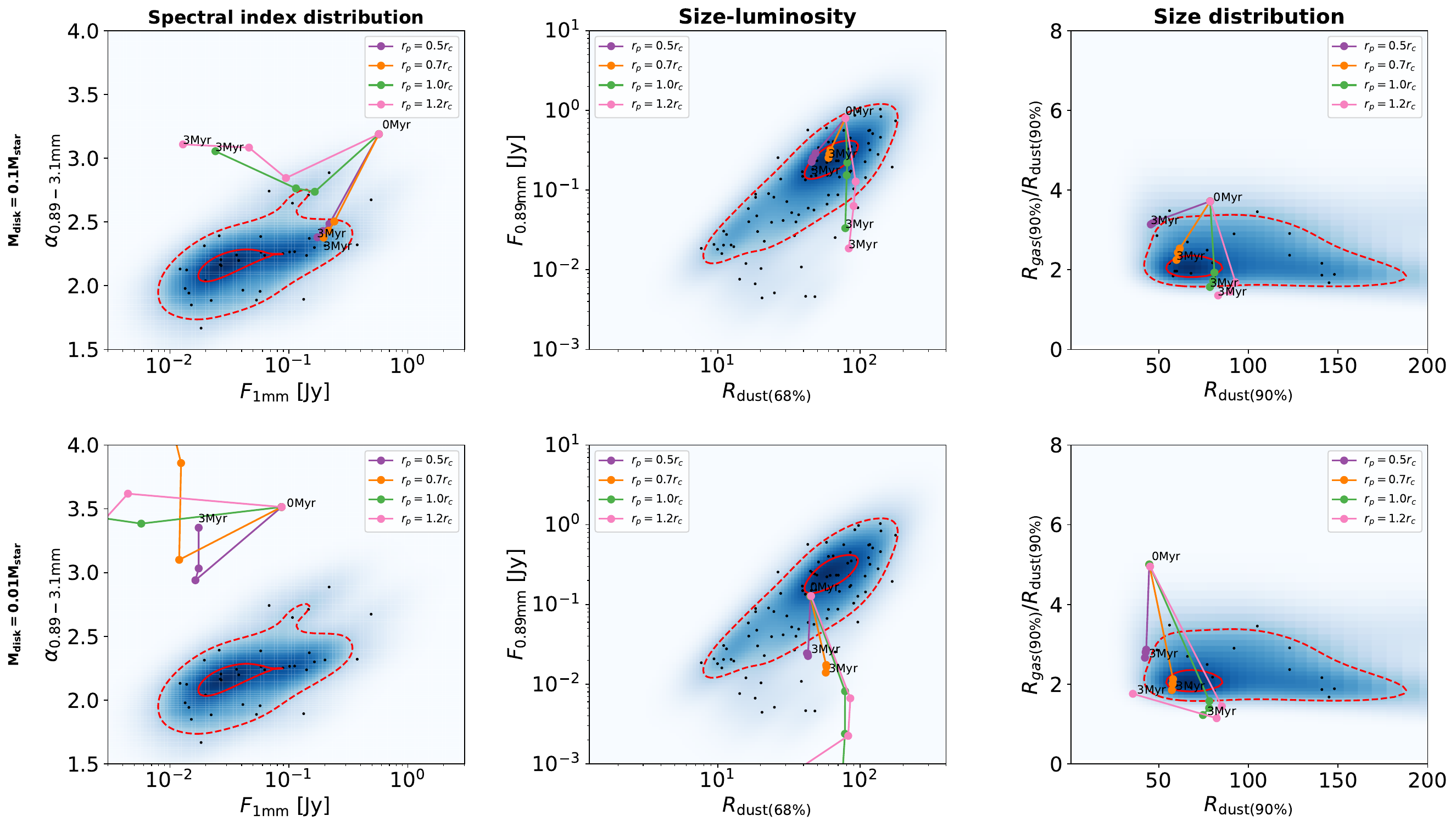}
    \caption{Evolution in the spectral index, size-luminosity, and $R_{\mathrm{gas(90\%)}}/R_{\mathrm{dust(90\%)}}$ vs $R_{\mathrm{dust(90\%)}}$ spaces of substructured disks evolved with \texttt{DustPy} code with external photoevaporation ($F_{\rm{FUV}} = \SI{4}{\mathrm{G}_{0}}$). Low viscosity regime ($\alpha=10^{-3.5}$) and disk characteristic radius fixed to $r_{c}=\SI{50}{au}$. Different values of the position of the inserted substructure $r_{p}$ have been explored. The points associated with each trajectory represent the snapshots taken at \SI{0}{Myr}, \SI{1}{Myr}, \SI{2}{Myr}, and \SI{3}{Myr}, respectively. Top row: $M_{\rm{disk}}=\SI{0.1}{\mathrm{M_{star}}}$. Bottom row: $M_{\rm{disk}}=\SI{0.01}{\mathrm{M_{star}}}$.}
    \label{fig:1substr_50rc}
\end{figure*}

\subsection{Possible solutions and future perspectives}

Our results show that including external photoevaporation does not fully resolve the discrepancy between the gas-to-dust size ratios of the disks predicted by the models and those observed in the Lupus population. Therefore, it may be necessary to reconsider the assumptions underlying the adopted disk evolution model. Could reducing the $\alpha$ viscosity parameter to lower values (e.g. $10^{-5}$), to mitigate viscous spreading, lead to the desired outcome? As shown in \cite{delussu2024population} and in Appendix \ref{appendix:lowest_alpha}, assuming such a low viscosity value results in the formation of disks with very low fluxes and overly small rings.

\begin{figure*}[thb]
    \centering
    \includegraphics[scale=0.45]{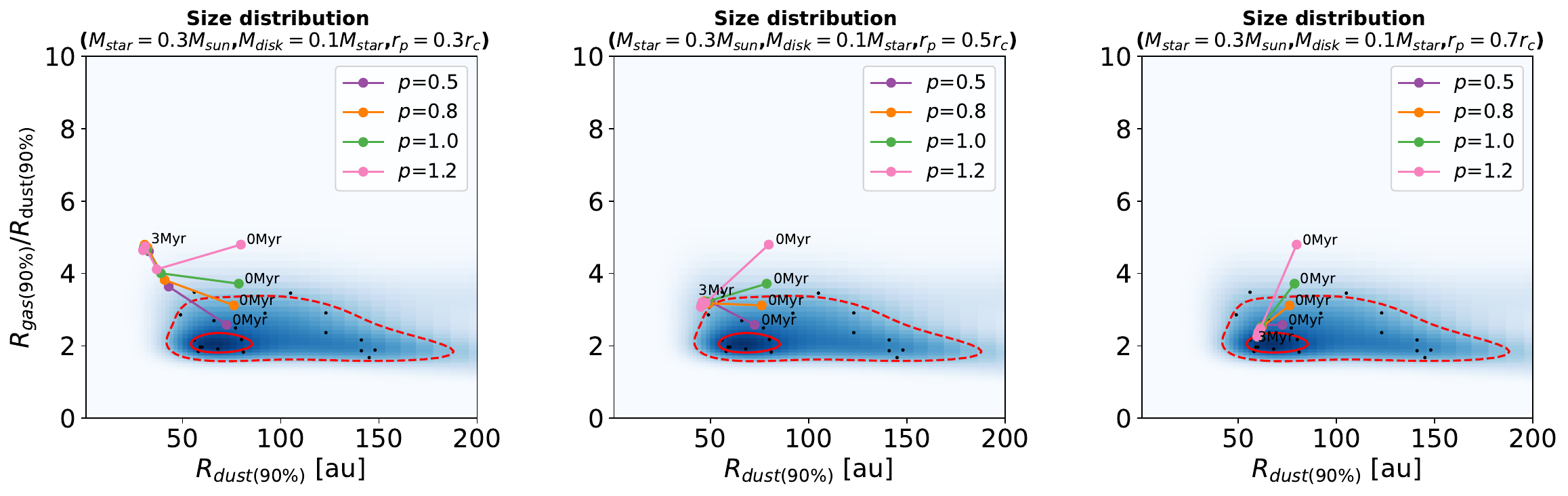}
    \caption{Evolution in the $R_{\mathrm{gas(90\%)}}/R_{\mathrm{dust(90\%)}}$ vs $R_{\mathrm{dust(90\%)}}$ space of substructured disks evolved with \texttt{DustPy} code with external photoevaporation ($F_{\rm{FUV}} = \SI{4}{\mathrm{G}_{0}}$). Low viscosity regime ($\alpha=10^{-3.5}$) and disk characteristic radius fixed to $\rc=\SI{50}{au}$. Different values of the position of the inserted substructure $r_{\rm{p}}$ and $p$ initial surface density power law exponent values have been explored. The points associated with each trajectory represent the snapshots taken at \SI{0}{Myr}, \SI{1}{Myr}, \SI{2}{Myr}, and \SI{3}{Myr}, respectively.}
    \label{fig:gamma_variation_rc50}
\end{figure*}

In addition, we considered a constant $\alpha$ value across the disk in this work. However, the dependence of $\alpha$ on the distance from the central star could alter disk’s evolution, including its radius over time. This radius-dependent $\alpha$ profile may, for example, arise due to the presence of dead zones within the disk \citep[e.g.,][]{tong2024question}.\\
\\
Viscous evolution, which leads the disk to spread to larger gas radii, might also be addressed by revisiting our assumption for the initial surface density power law exponent $p$ (Eq. \ref{eq:surface_density}). Indeed, as shown in \cite{toci2021secular}, $p$ (in their notation $\gamma$) has a strong impact on both dust and gas radii. They highlight that gas radii are significantly affected by the exact shape of the outer part of the disk which is ultimately determined by $p$. Lower values of $p$ imply a sharper outer edge, leading to smaller gas radii. Nevertheless, they also show that the effect on dust radii of the choice of $p$ is mitigated after 2-3 Myr due to radial drift. Fig. \ref{fig:gamma_variation_rc50} shows that, in the case of combined substructures and external photoevaporation, varying $p$ results in evolutionary tracks that rapidly converge towards similar behaviors. This occurs because changing $p$ in the initial gas density profile, but adopting a radially constant $\alpha$ leads to an out-of-equilibrium condition. As a result, disks tend to go back to the $p=1$ configuration after a few viscous timescales. However, Fig. \ref{fig:gamma_variation_rc50} shows a promising result as reducing $p$ reduces the initial gas-to-dust size ratio due to the expected reduction in gas radii for $p<1$.
Appendix \ref{appendix:gamma_alpha_var} presents the result obtained by combining external photoevaporation and the variation of $p$ with a radially decreasing $\alpha$ profile, which would avoid disks from falling back to the $p=1$ scenario. Additionally, this scenario would produce flatter radial profiles of the maximum grain size, in agreement with what has been recently observed by multi-wavelength studies \citep[e.g,][]{sierra2021molecules,guidi2022distribution,jiang2024grain}. Nevertheless, the result presented in Appendix \ref{appendix:gamma_alpha_var} shows disk evolution similar to the case presented in Fig. \ref{fig:gamma_variation_rc50}. This is due to the fact that external photoevaporation dominates the shaping of the outer disk region, which ultimately sets the final disk radii. As a result, although the combination of a radially decreasing $\alpha$ and variation in $p$ modifies the overall surface density profile, the impact on the outer disk is effectively erased by external photoevaporation. What remains is a difference in the inner slope, which, as discussed above, does not significantly impact the final disk sizes. This outcome mirrors our previous result when varying only $p$, confirming that external photoevaporation is the primary driver of disk sizes, and explains that the difference with \cite{toci2021secular} arises from the inclusion of external photoevaporation in our model.

\begin{figure*}[thb]
    \centering
    \includegraphics[scale=0.4]{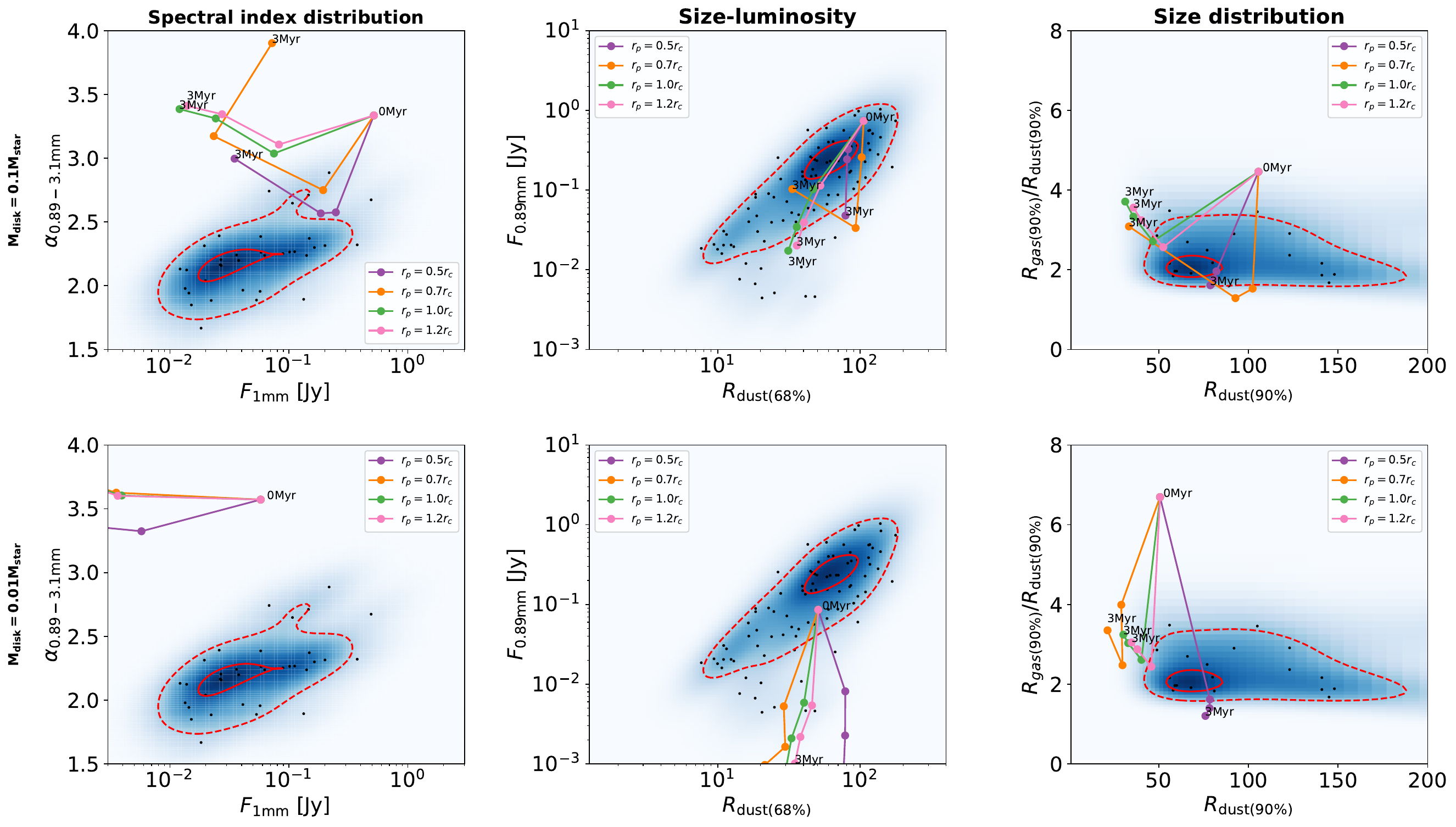}
    \caption{Same as Figure \ref{fig:1substr_50rc}, but disks with characteristic radius fixed to $\rc=\SI{100}{au}$.}
    \label{fig:1substr_100rc}
\end{figure*}

An alternative option is to consider a different scenario, where disk evolution is driven by magnetohydrodynamics (MHD) winds. In contrast to purely viscous evolution, where angular momentum is radially transported across the disk, allowing accretion onto the star, MHD winds extract material (and angular momentum) from the disk. This last mechanism drives disk evolution and accretion without spreading the disk, possibly keeping gas disk sizes small enough to fulfill observational constraints. The discrepancy between the observed and predicted gas and dust disk radii may offer a way to constrain the relative contributions of viscous- and wind-like mechanisms.  To compare the two disk evolution frameworks, the implementation of an MHD-wind parametrization \citep[e.g.,][]{Tabone2022} in the numerical codes used is needed and still to be tested.\\
\\
In this study, substructures are modeled as gaps associated with the presence of a planet inserted in the disk. These substructures are assumed to be static, and planetary migration is therefore not included. Although a planet-based prescription is adopted, the analysis is not intended to trace the physical origin or evolutionary history of the substructures. Instead, we adopt an agnostic perspective regarding their origin, treating substructures as a generic representation of pressure maxima in disks rather than uniquely ascribing them to migrating planets.\\ If the substructures considered in this work are interpreted as planet-induced gaps, migration effects may become relevant. However, for the low-viscosity disks and strong dust trapping regimes considered here, gap-opening planets are expected to migrate in the Type II regime, with migration proceeding on the viscous timescale or slower. Under these conditions, the inward displacement of the pressure maximum is expected to remain modest over the disk lifetimes considered and is therefore unlikely to significantly affect the disk radii, which constitute the main observables of this study. A more comprehensive and quantitative assessment of migration effects will be explored in future work.\\
\\
To further enhance the model’s completeness, we plan to include internal photoevaporation in future work to assess its effects quantitatively. Previous theoretical studies combining internal and external photoevaporative winds have shown that disks can follow different dispersal pathways depending on the relative strength of these processes, with internal photoevaporation dominating in some regimes and driving inside-out dispersal in others \citep{coleman2022dispersal}. For the disk population considered here, internal photoevaporation is expected to primarily affect disks at late evolutionary stages, once accretion rates have dropped to low values. In this regime, its main effect is likely to be the rapid clearing of the remaining gas and the termination of disk evolution, rather than a significant modification of disk structure over Myr timescales. Indeed, external processes, such as external photoevaporation, are expected to dominate the early evolution of large disks \citep{clarke2001dispersal}. A more detailed treatment of internal photoevaporation will be explored in future work.\\
\\
Both the stellar luminosity, $L_{\star}$, and the effective temperature, $T$, are not evolved in our simulations. As shown in \citet{delussu2024population}, the results obtained when $L_{\star}$, and thus $T$ are allowed to evolve are analogous to those from the fixed-luminosity scenario. Furthermore, while the evolution of stellar luminosity is expected to have a stronger impact for more massive stars ($\gtrsim1.5\,\rm{M_{\odot}}$) \citep[e.g.,][]{kunitomo2021photoevaporative,ronco2024planet}, this effect is limited in our study because massive stars do not significantly contribute to our disk population. Indeed, for the two-pop-py simulations, stellar masses have been randomly sampled from the IMF to reflect the distribution observed in star-forming regions, while for the dustpy simulations, we adopted two representative stellar masses ($0.3\, \rm{M_\odot}$ and $1\, \rm{M_\odot}$).\\
\\
Gravitational instability (GI) can play an important role in massive disks, potentially affecting their structure, dust evolution, and observational signatures at early evolutionary stages. In the disk population adopted in the two-pop-py study, roughly 27\% of disks have high disk-to-star mass ratios ($M_{\rm{disk}}/M_{\rm{star}} \geq 0.2$), making them potentially prone to gravitational instability. Removing this subset produces only modest changes.
The mean millimeter continuum flux decreases by approximately $30-40\%$ across the different disk populations, leading to small shifts in the spectral index and size-luminosity distributions.
Gas and dust disk radii show an average decrease of less than $\sim10\%$, and their distributions remain qualitatively unchanged. Given that gas and dust disk radii are the main observables considered in this study, the inclusion of potentially unstable disks does not affect our population-level conclusions. The effects of gravitational instability will be addressed in a follow-up study, where a self-consistent treatment will be implemented.

\section{Conclusions}\label{conclusions}

In this work, we conducted a study to determine whether substructured disks can reproduce the observed gas-to-dust size ratio of the protoplanetary disk population observed in the Lupus star-forming region. Firstly, we performed a population synthesis study of substructured disks using the two-pop-py 1D evolutionary model for dust and gas in protoplanetary disks.
Subsequently, we explored the dust-gas size behavior of disks exposed to a moderate environmental FUV radiation field of \SI{4}{\mathrm{G}_{0}}, as the average registered in the Lupus region. The last investigation was performed using the \texttt{DustPy} code extended by an external module that includes the effect of an external FUV field. We compared our simulated disks to the observed dust-gas size distribution of disks of the Lupus star-forming region presented in \cite{sanchis2021measuring}. These are the main results we have outlined:
\begin{enumerate}
    \item We confirmed that smooth disks produce a larger $R_{\mathrm{gas(90\%)}}/R_{\mathrm{dust(90\%)}}$ compared to observations (Fig. \ref{fig:smooth_vs_substr}), extending \cite{toci2021secular} result to the broader level of a disk population synthesis.
    \item We showed, on a population synthesis level, that, as suggested by \cite{toci2021secular}, substructured disks, whether with one or two substructures, produce lower $R_{\mathrm{gas(90\%)}}/R_{\mathrm{dust(90\%)}}$ compared to smooth disks (Fig. \ref{fig:smooth_vs_substr}). However, although the presence of substructures helps mitigate the discrepancy between simulation and observation, it does not completely resolve it.
    \item We identified a tension for substructured disks between the favorable results in \cite{delussu2024population}, which show that the spectral index and size-luminosity distributions can be reproduced by simple constraints on their initial conditions, and the difficulty in matching the observed gas-to-dust ratios (Fig. \ref{fig:smooth_vs_substr}), even when accounting for external photoevaporation (Fig. \ref{fig:dustpy_lowviscosity}, \ref{fig:1substr_50rc}, \ref{fig:1substr_100rc}).
    \item Specific combinations of initial conditions can produce results consistent with observations for disks undergoing viscous evolution with external photoevaporation (Fig. \ref{fig:1substr_50rc}). However, the restricted range of initial conditions needed for these outcomes introduces a warning for a fine-tuning problem.
    \item Both population studies exposed the difficulty in reproducing the observed gas-to-dust size ratios. The presence of substructures can reproduce the spectral index and dust size but leads to an overestimate of gas radii. On the other hand, when gas radii are reproduced as in \cite{toci2021secular}, dust radii are underestimated. This ultimately emphasizes that the main issue lies in simultaneously reproducing both gas and dust sizes. One possible explanation is that the outermost substructure is linked to the disk truncation radius, which ultimately determines the gas radius, or that substructures are so frequent within each disk that one is always found near the gas outer radius. This may point towards an inability of viscous evolution to reproduce observations if paired with constraints of dust evolution.
\end{enumerate}
In future work, we aim to extend our investigation by broadening the parameter space of the initial conditions and exploring further scenarios such as wind-driven disks, GI, internal photoevaporation, and planetary migration.

\begin{acknowledgements}
    L.D., T.B., T.C.H.L., and S.M.S. acknowledge funding from the European Union under the European Union's Horizon Europe Research and Innovation Programme 101124282 (EARLYBIRD) and funding by the Deutsche Forschungsgemeinschaft (DFG, German Research Foundation) under grant 325594231, and Germany's Excellence Strategy - EXC-2094 - 390783311.\\
    G.R. and R.A. acknowledge funding from the Fondazione Cariplo, grant no. 2022-1217, and the European Research Council (ERC) under the European Union’s Horizon Europe Research \& Innovation Programme under grant agreement no. 101039651 (DiscEvol).\\
    Views and opinions expressed are, however, those of the authors only and do not necessarily reflect those of the European Union or the European Research Council. Neither the European Union nor the granting authority can be held responsible for them.
\end{acknowledgements}

\bibliographystyle{aa}
\bibliography{biblio2}

\onecolumn
\begin{appendix}
    \section{Dustpy vs two-pop-py}\label{appendix:dustpy_vs_twopoppy}
\begin{figure}[H]
        \centering
        \includegraphics[scale=0.4]{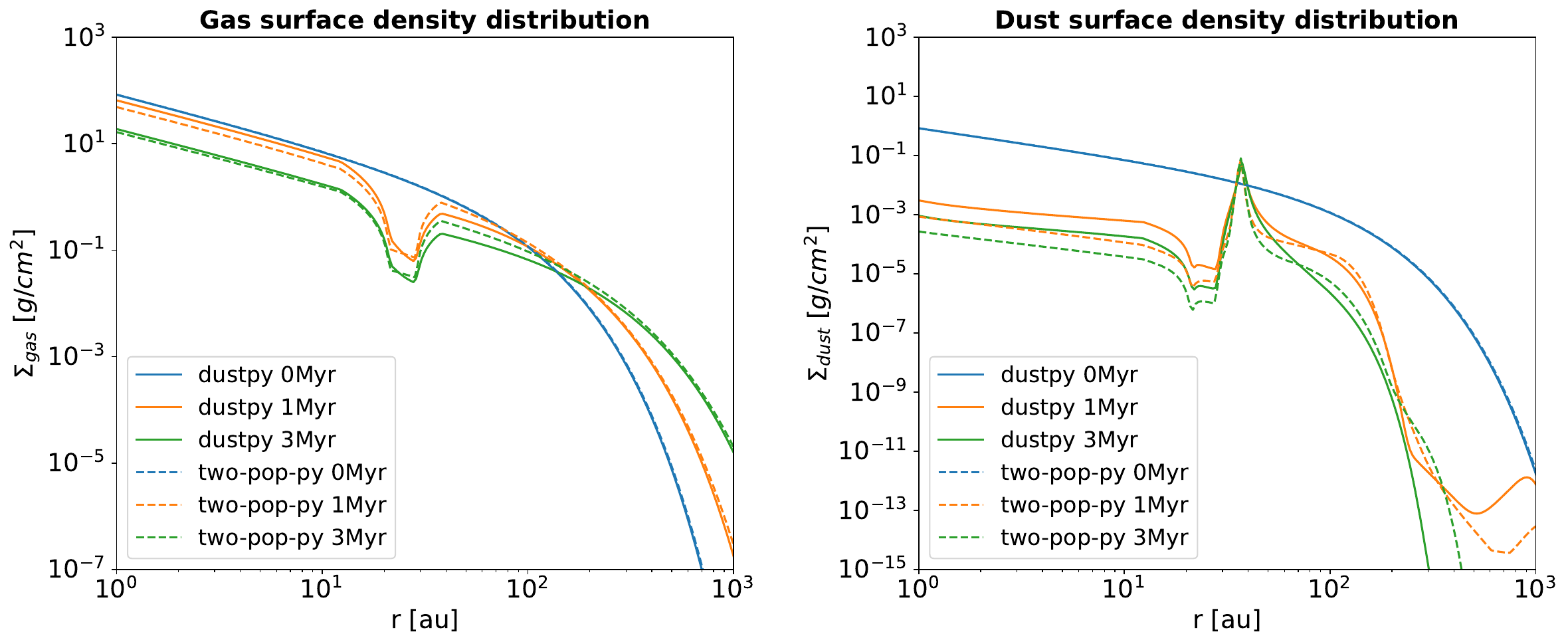}
        \caption{Comparison of gas surface density (left) and dust surface density (right) evolution between dustpy and two-pop-py models at \SI{0}{Myr}, \SI{1}{Myr}, and \SI{3}{Myr} for a substructured test disk with: $\alpha = 0.001$, $M_{\rm{star}} = \SI{0.3}{M_{\rm{sun}}}$, $M_{\rm{disk}} = \SI{0.01}{M_{\rm{star}}}$,  $r_{c}=\SI{50}{au}$, $v_{frag}=\SI{1000}{cm/s}$, $r_{p}=\SI{0.5}{r_{c}}$, and $m_{\rm{p}} = \SI{1}{M_{\rm{J}}}$.}
        \label{fig:surface_dustpy_fraglim_vs_twopoppy}
    \end{figure}

\begin{figure}[H]
        \centering
        \includegraphics[scale=0.45]{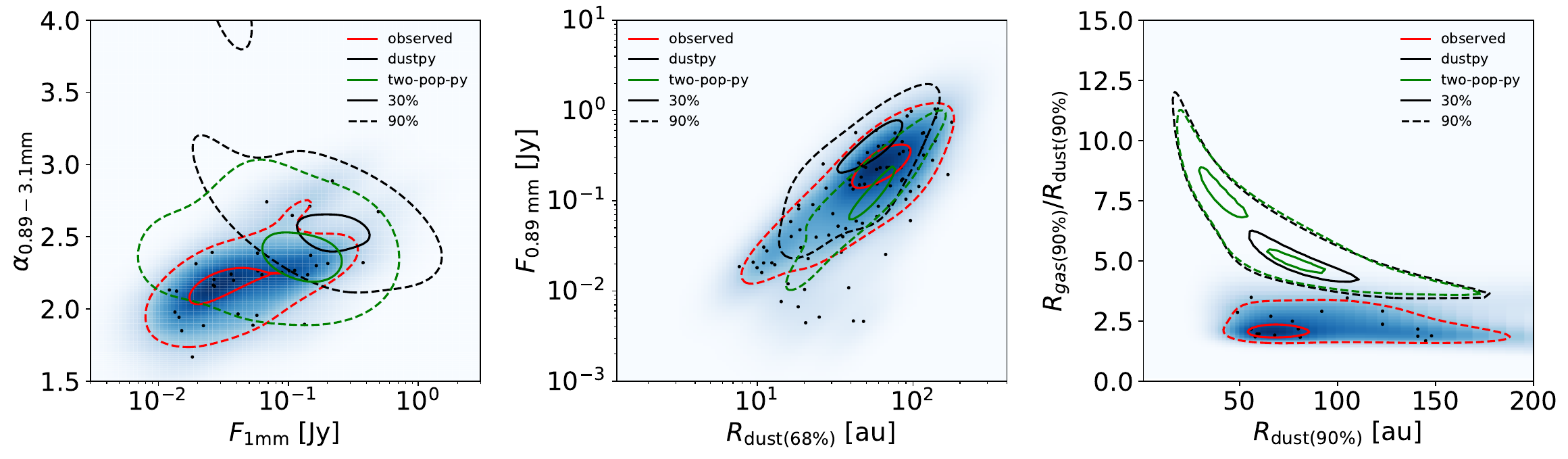}
        \caption{Spectral index distribution (left), size-luminosity distribution (middle), and gas-dust size distribution (right) comparison between \texttt{DustPy} and two-pop-py at 1 Myr for a test population of substructured disks. Both test populations are constructed by sampling all combinations of the following parameters: $\alpha = 0.001$, $M_{\rm{star}} = [ 0.3, 1.0]\ \rm{M_{sun}}$, $M_{\rm{disk}} = \SI{0.01}{M_{\rm{star}}}$,  $r_{c}=[20, 50, 100]\ \rm{au}$, $v_{frag}=\SI{1000}{cm/s}$, $r_{p}=[ 0.5, 0.7]\ \rm{r_{c}}$, and $m_{\rm{p}} = \SI{1}{M_{\rm{J}}}$.}
        \label{fig:observables_dustpy_fraglim_vs_twopoppy}
    \end{figure}

    Appendix \ref{appendix:dustpy_vs_twopoppy} shows a comparison between \texttt{DustPy} and two-pop-py for substructured test disks. We show the evolution of gas and dust surface densities (Fig.~\ref{fig:surface_dustpy_fraglim_vs_twopoppy}) for a substructured test disk of initial disk mass $M_{\rm{disk}}= \SI{0.01}{\mathrm{M_{star}}}$, initial characteristic radius $r_{\rm{c}}= \SI{50}{\mathrm{au}}$, $\alpha = 0.001$, $v_{frag}=\SI{1000}{cm/s}$, around a star of mass $M_{\rm{star}} = \SI{0.3}{M_{\rm{sun}}}$, and hosting a planet of mass $m_{\rm{p}} = \SI{1}{M_{\rm{J}}}$ at $r_{p}=\SI{0.5}{r_{c}}$.
    In Fig. ~\ref{fig:observables_dustpy_fraglim_vs_twopoppy}, we show the disk radii, size-luminosity, and spectral index distributions for a population of test disks with initial conditions as listed in the caption of the Figure. The models produce similar disk evolution and properties: disk radii and size–luminosity distributions are in good agreement, while modest differences appear near the planet-induced substructure, where two-pop-py produces slightly stronger dust trapping and a sharper pressure maximum. These local variations lead to overlapping but not precisely matching spectral index distributions. However, they do not change the disk radii, which are the main observable considered in this study, and therefore do not affect the gas-to-dust size distributions. Since two-pop-py was not designed or calibrated for substructure, and \texttt{DustPy} includes a more detailed treatment of gap formation, \texttt{DustPy} is adopted for the modeling presented in this study.
    \section{Low viscosity case}\label{appendix:lowest_alpha}
\begin{figure}[H]
        \centering
        \includegraphics[scale=0.4]{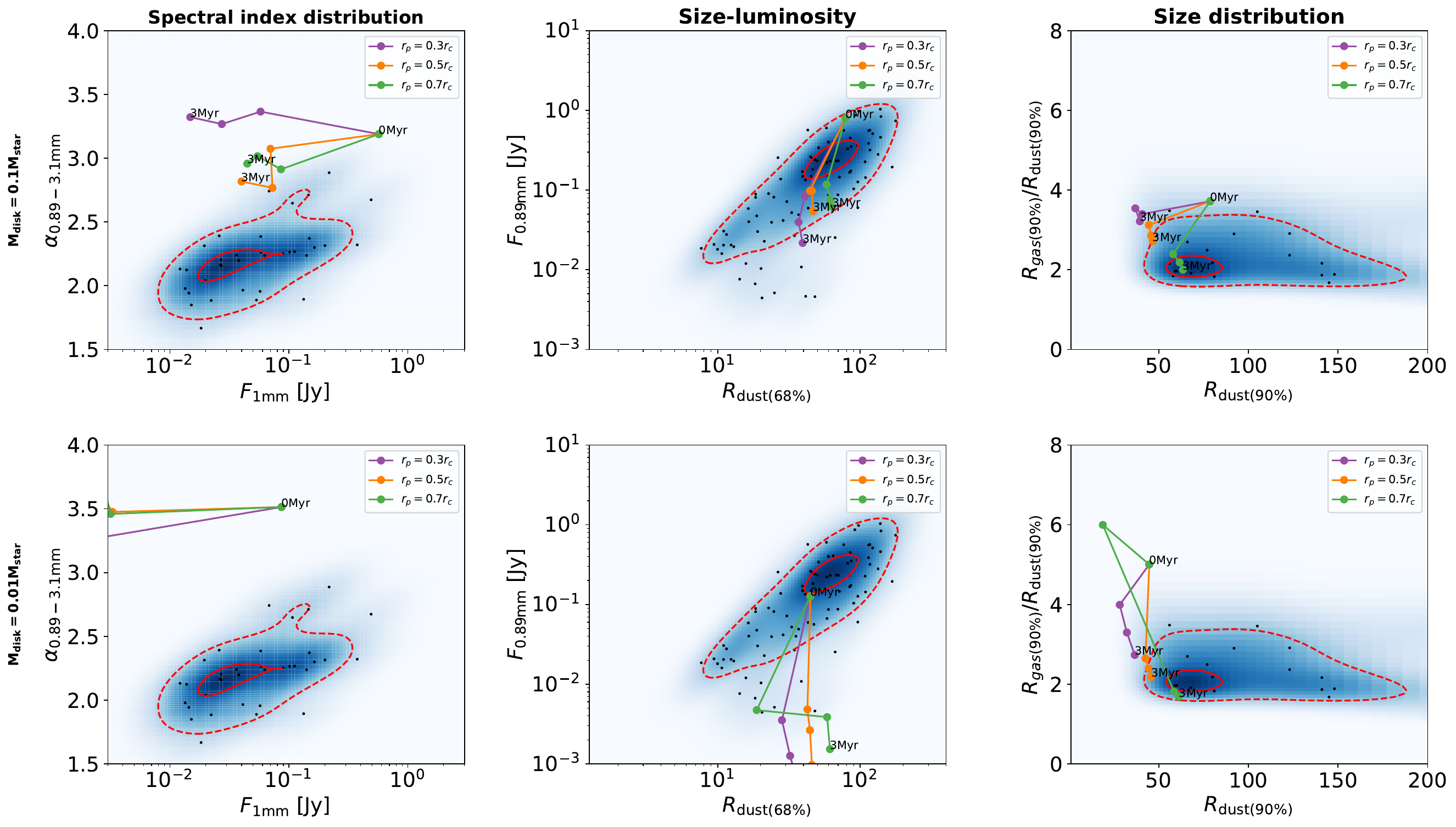}
        \caption{Same as Figure \ref{fig:1substr_50rc}, but disks with $\alpha=10^{-5}$.}
        \label{fig:lowest_alpha}
    \end{figure}

    In this Appendix, we present the results obtained for a very low $\alpha$ viscosity regime (i.e., $\alpha=10^{-5}$), investigating whether reducing the viscosity could help mitigate viscous spreading in the disk. As shown in \cite{delussu2024population} and in Fig. \ref{fig:lowest_alpha}, however, adopting such a low viscosity leads to the formation of disks with very low fluxes and overly small rings, which do not match the observed population.
    \section{Varying the disk surface density profile and the disk viscosity}\label{appendix:gamma_alpha_var}
\begin{figure}[H]
        \centering
        \includegraphics[scale=0.45]{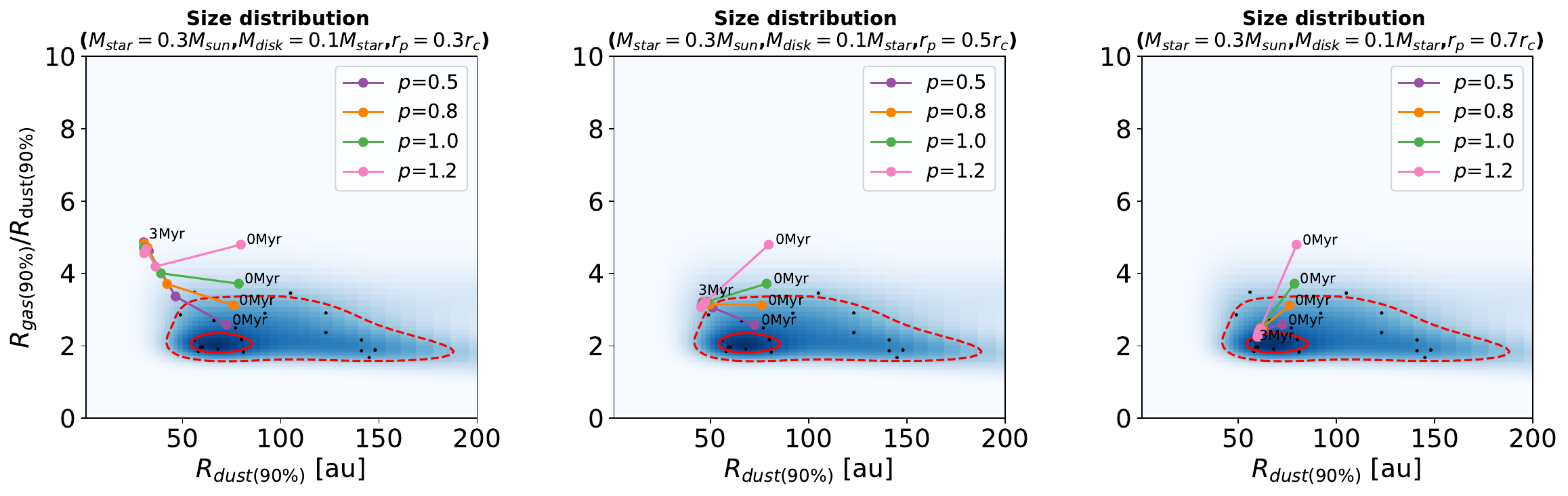}
        \caption{Evolution in the $R_{\mathrm{gas(90\%)}}/R_{\mathrm{dust(90\%)}}$ vs $R_{\mathrm{dust(90\%)}}$ space of substructured disks evolved with \texttt{DustPy} code with external photoevaporation ($F_{\rm{FUV}} = \SI{4}{\mathrm{G}_{0}}$). Disk characteristic radius fixed to $\rc=\SI{50}{au}$. Different values of the position of the inserted substructure $r_{\rm{p}}$ and $p$ initial surface density power law exponent values have been explored. The choice of $p$ has been combined with a radially decreasing $\alpha$ profile: $\alpha = \alpha_{0} (r/r_{c})^{-p-1}$ with $\alpha_{0}=10^{-3.5}$. The points associated with each trajectory represent the snapshots taken at \SI{0}{Myr}, \SI{1}{Myr}, \SI{2}{Myr}, and \SI{3}{Myr}, respectively.}
        \label{fig:gamma_alpha_app}
    \end{figure}

    In this Appendix, Fig. \ref{fig:gamma_alpha_app} shows the evolution of $\rm{R_{gas}}/\rm{R_{dust}}$ as a function of $\rm{R_{dust}}$ under external photoevaporation, varying the surface density density profile $p$ (Eq. \ref{eq:surface_density}), and allowing the $\alpha$ viscosity parameter to decrease with distance to the central star: $\alpha = \alpha_{0} (r/r_{c})^{-p-1}$, with $\alpha_{0}=10^{-3.5}$.
\end{appendix}
\end{document}